\documentclass[a4paper,fleqn,usenatbib,letter]{mnras}

\usepackage{mathptmx}
\usepackage[T1]{fontenc}
\usepackage{ae,aecompl}
\usepackage{graphicx}        
\usepackage[dvipsnames]{xcolor}
\usepackage{amsmath}
\usepackage{tabularx}
\usepackage{arydshln}
\usepackage{float}

\def\mnras{MNRAS}
\def\ApJ{Astrophysical Journal}
\def\apjl{\em Ap. J. Lett.}
\def\CQG{Class. Quantum Grav. }
\def\PRD{Phys. Rev. D}

\def\pertp{\varepsilon}

\def \Sb {\Sigma_0}
\def \ro {a}
\def \rf {r_+}

\def \nuf {\nu^+}
\def \lf {\lambda^+}
\def \of {\tilde{\omega}^+}
\def \mf {m_0^+}

\def \Mf {M^+}

\def \htwof {h_2^+}
\def \vf {\upsilon^+}

\def \rv {r_-}

\def \nuv {\nu^-}
\def \lv {\lambda^-}
\def \ov {\tilde{\omega}^-}
\def \mv {m_0^-}
\def \hv {h_0^-}

\def \htwov {h_2^-}

\def\fpt{K_1}
\def\spt{K_2}

\def\energy{E}
\def\pressure{P}
\def\Eb{\energy}

\def\Pb{\pressure}

\def\Ppp{\pressure^{(2)}}

\def \Etppz{\tilde{\energy}^{(2)}_0}
\def \Ptpp{\tilde{\pressure}^{(2)}}
\def \Ptppt{\tilde{\pressure}^{(2)}_2}
\def \Ptppz{\tilde{\pressure}^{(2)}_0}

\def \radius{R_T}

\def\Qt2{\hat{Q}_2}

\newcounter{mnotecount}[section]

\renewcommand{\themnotecount}{\thesection.\arabic{mnotecount}}
\newcommand{\mnote}[1]%{}
{\protect{\stepcounter{mnotecount}}$^{\mbox{\footnotesize
$%\!\!\!\!\!\!\,
\bullet$\themnotecount}}$ \marginpar{%\color{red}%
\raggedright\tiny\em
$\!\!\!\!\!\!\,\bullet$\themnotecount: #1} }

\title[Completion of the universal $I$-Love-$Q$ relations in compact stars including the mass]{Completion of the universal $I$-Love-$Q$ relations in compact stars including the mass}

\author[B.~Reina et al.]{
Borja Reina,$^{1,2}$\thanks{ 
E-mail: borja.reina@dcu.ie}
Nicolas Sanchis-Gual,$^{3}$
Ra\"ul Vera$^{1}$
and Jos\'e A.~Font$^{3,4}$
\\
$^{1}$Departamento de F\'isica Te\'orica e Historia de la Ciencia, University of the Basque Country UPV/EHU, Apartado 644, 48080 Bilbao, Spain\\
$^{2}$School of Mathematical Sciences,  Dublin City University, Glasnevin,  Dublin 9,  Ireland \\
$^{3}$Departamento  de  Astronom\'ia  y  Astrof\'isica,  Universitat  de  Val\`encia,
Dr.   Moliner  50,  46100,  Burjassot  (Val\`encia),  Spain\\
$^{4}$Observatori Astron\`omic, Universitat de Val\`encia, C/ Catedr\'atico 
  Jos\'e Beltr\'an 2, 46980, Paterna (Val\`encia), Spain
}

\date{Accepted XXX. Received YYY; in original form ZZZ}

\pubyear{2017}

\begin{document}
\label{firstpage}
\pagerange{\pageref{firstpage}--\pageref{lastpage}}
\maketitle

\begin{abstract}
In a recent paper we applied a rigorous perturbed matching framework to show the amendment of the mass of rotating stars in Hartle's model. Here, we apply this framework to the tidal problem in binary systems. Our approach fully accounts for the correction to the Love numbers needed to obtain the  universal $I$-Love-$Q$ relations. We compute the corrected mass vs radius configurations of rotating quark stars, revisiting a classical paper on the subject. These corrections allow us to find a universal relation involving the second-order contribution to the mass $\delta M$. We thus complete the set of universal relations for the tidal problem in binary systems, involving four perturbation parameters, namely $I$, Love, $Q$, {\it and} $\delta M$. These relations can be used to obtain the perturbation parameters directly from observational data.
\end{abstract}


%%%%%%%%%%%
\section{Introduction}
%%%%%%%%%%%

The construction of analytic models of astrophysical compact bodies in general relativity (GR) relies on the matching of spacetimes theory. The  idea is to consider two different bounded regions, namely an interior fluid and a vacuum exterior, and to impose appropriate matching conditions on a timelike hypersurface $\Sigma$ separating them. Therefore, a global model is constructed by joining the common boundary data on $\Sigma$. While the search for an exact global model for a rotating compact body is a major challenge, the situation becomes tractable when one resorts to approximate methods such as perturbation theory. In this context, models that describe rotating stars \citep{Hartle1967}, tidal effects \citep{DamourNagar}, or collapsing stars \citep{Brizuela} have been developed.

Although the matching of spacetimes in the exact case has been well understood for decades, 
the matching in a perturbative scheme has needed a much longer concoction. The first fully general
and consistent perturbation theory of hypersurfaces to second order is due to~\citet{Mars2005}. 
On top of a background spacetime where two regions are matched on some $\Sb$, first and second-order 
problems for the corresponding regions are developed. The theory provides the common boundary data 
on $\Sb$ for such problems and the \emph{gauge-independent} equations for the quantities that describe 
the deformation of the surface. Perturbed matching is commonly treated in the literature by 
prescribing some extension of the exact matching conditions to the perturbative scheme, or assuming 
the continuity  of the functions driving the perturbations across $\Sb$ (see \cite{MaMeVe2007}). However, 
this is not ensured a priori, and assuming explicit choices of coordinates (and gauges) in which the perturbations satisfy certain continuity and differentiability conditions may subtract generality to the model.
Even worse, it may lead to wrong outcomes. We discuss two relevant examples next.

The first example has to do with slowly-rotating relativistic stars. In their pioneering work, Hartle and Thorne presented the general relativistic treatment of isolated rotating compact stars in 
equilibrium~\citep{Hartle1967, HartleThorne1968}, known as ``Hartle's model'' in short. This stands as the 
basis to construct analytical models in axial symmetry (see \citet{LivRev_Stergioulas} and references therein).  A perturbative scheme is built upon a spherical non-rotating configuration, on top of which stationary (rotating) and axial perturbations are taken to second order. Models are built assuming a perfect fluid interior with barotropic equation of state (EOS) that rotates rigidly and with no convective motions. Under these assumptions the perturbations are described by four functions, whose values at the surface of the star determine the dragging of inertial frames, the deformation, and the total mass of the star in terms of the central density. At each order those values are computed (i) integrating from the regular centre given an EOS, (ii) solving the asymptotically flat vacuum exterior, and (iii) assuming the continuity of the functions across the surface choosing an explicit coordinate system.  

Assumption (iii) has significant implications on the properties of the stellar models. As shown 
by~\citet{ReinaVera2014} using the framework of~\citet{Mars2005}, a relevant perturbative function 
presents a discontinuity proportional to the (background) energy density in the stellar surface. This function enters the computation of the mass of the star at second order $\delta M$ and, therefore, the original expression for the total mass of the rotating star given in~\cite{Hartle1967} has to be amended. Idealised, constant-density stars, originally studied by~\cite{Chandra-Miller1974}, were subsequently analyzed in~\cite{BorjaHomogeneous}, showing that the deviations in the mass-radius diagrams are far from negligible.
%This issue is also present in the Newtonian case \citep{Reina16}.
 
The second example has to do with the so-called $I$-Love-$Q$ relations
\citep{YagiYunes}, where ``$I$'' and ``$Q$'' refer to the moment of inertia and the 
quadrupole moment of the star respectively, and the Love numbers ``$k_l$'' are
associated with the tidal field due to the presence of a companion
star. The most basic treatment of this problem fits in Hartle's scheme
and can be solved in the regime of stationary and axial perturbations (see
\cite{DamourNagar} and references therein).  Although $I$, $k_l$ and
$Q$ depend individually strongly on the EOS, they are related in an
EOS-independent way. \cite{YagiYunes} found that these relations split
into two different categories: ordinary EOS stars and  quark stars.
However, using a correction identified in \cite{DamourNagar} as a
pathological behaviour of one of the field equations across the
surface of homogeneous stars,~\citet{YagiYunes_erratum} obtained
an amended  expression for the Love numbers
that leads to universal $I$-Love-$Q$ relations.
As hinted by the correction to the original Hartle's model,
it is straighforward to show that the solution to the tidal problem
is indeed related: the result of~\cite{YagiYunes} was incorrect due to the
assumption of continuity of the perturbations across $\Sb$,
because a function connected to the Love numbers exhibits a
discontinuity proportional to the energy density at the stellar 
surface, which does not vanish in quark stars. Let us note that the jumps
in the tidal problem were already addressed by \cite{ThornePrice} and \cite{CampolattaroThorne70}.
Although exact only for $l\geq 2$, these discontinuities had somehow been forgotten.

The aim of this {\it Letter} is to recall the perturbed matching conditions from~\cite{ReinaVera2014}
to then: (i) apply the corrections in the context of Hartle's model to quark stars with linear EOS, revisiting 
the work of~\cite{ColpiMiller}; (ii) set forth the discontinuities involved in the $l\geq 2$ sector of the 
first-order tidal-field problem and show how those conditions, which indeed coincide with those 
in~\cite{ThornePrice}, lead to the corrections recognised in~\cite{DamourNagar} and used 
in~\cite{Hinderer_tidal_waves} and~\cite{YagiYunes_erratum}; (iii) complete the $I$-Love-$Q$ relations
incorporating $\delta M$, now correctly computed, that indicate universal relations for all perturbative 
quantites.
% Let us remark that our corrections 
% are derived from a consistent, general, EOS-independent, perturbed matching theory.
% The discontinuities in a most general setting can be extracted from the geometrical
% perturbed matching conditions found in \cite{ReinaVera2014}. In GR
% these will be proportional to the jump of the energy density at $\Sb$ 
% and could be used to construct realistic models of compact stars.

%%%%%%%%%%%%%%%%%%%
\section{Setting and field equations}
\label{setting}
%%%%%%%%%%%%%%%%%%%

We start with the construction of the global background configuration. Consider two static spherically symmetric spacetimes $(\mathcal{V}^+, g^+, \Sigma_0^+)$ and $(\mathcal{V}^-, g^-, \Sigma_0^-)$, with 
$g^\pm = -e^{\nu^\pm} dt_\pm^2 + e^{\lambda^\pm} dr_\pm^2 + r_\pm^2 (d\theta_\pm^2 + \sin^2 \theta_\pm d \varphi_\pm^2)$, that are matched across the boundaries, $\Sb :=\Sigma_0^\pm = \{r^\pm = \ro^\pm\}$ with constants $\ro^\pm>0$. The interior region $(+)$ radial coordinate ranges in $r^+ \in (0, \ro^+)$ and the exterior $(-)$ in $r^- \in (\ro^-, \infty)$. The matching conditions for this setting are $\ro := \ro^+ = \ro^-, \, [\lambda] = [\nu ] = [\nu' ] =   0$, where $':=d/dr$ and  $[f]$ is the difference of the function $f$ evaluated on $\Sigma_0$ from the $(+)$ and $(-)$ sides, i.e. $[f] := f^{+}|_{\Sb} - f^{-}|_{\Sb}$. 

The perfect fluid interior is described by its unit fluid flow $\vec{u}$, whose energy density $E\geq 0$ and pressure $P\geq 0$ are related by a barotropic EOS $E(P)$. The mass function is defined by $e^{-\lambda+(\rf)} = 1-2M^+(\rf)/\rf$. The TOV equations hold \citep{Hartle1967} and determine the interior configuration given the central value of the energy density $E(0):=E_c$. Function $\nu$ is determined up to an additive constant. The asymptotically flat vacuum exterior is Schwarzschild, determined by the total mass $M$, explicitly $e^{\nuv(\rv)} = e^{-\lv (\rv)} = 1-\frac{2M}{\rv}$.

Given this matter content, the matching conditions are interpreted as follows: $[\lambda] = 0$ fixes the constant $M$ to be the mass of the fluid, $M=\Mf(\ro)$, $[\nu]=0$ fixes the value of $\nu^+$ at the origin, and $[\nu']=0$ is just $P(\ro)=0$, which determines $\ro$. 

%%%%%%%%%%%%%%%%%%%%%%%%%%
\subsection{Perturbative scheme for rotating stars}
%%%%%%%%%%%%%%%%%%%%%%%%%%

Hartle's model is based upon the following forms for the
perturbation tensors in each region (let us drop $\pm$ for clarity) to first and second order respectively\footnote{
The function $m(r,\theta)$ in Eq.~(\ref{sopert_tensor}) corresponds to the $m$ used in \cite{Hartle1967}, whereas $m$ in \cite{ReinaVera2014} is $r e^{-\lambda}m$ here.
}
\begin{align}
  \fpt^{H}  =& 
               -2r^2\, \omega(r) \sin ^2 \theta dt d\varphi \label{fopert_tensor},\\
  \spt^{H} =& \left(-4 e^{\nu(r)} h(r, \theta) + 2r^2 \sin ^2 \theta {\omega}^2(r)\right)dt^2 \label{sopert_tensor}\\
            & + 4 e^{2\lambda(r)}r^{-1} m(r, \theta) dr^2 +4 r^2 k(r, \theta)
              (d\theta^2 + \sin ^2 \theta d\varphi^2),\nonumber
\end{align}
where $h(r,\theta) = h_0(r) + h_2(r) P_2(\cos \theta)$, $m(r,\theta) = m_0(r) + m_2(r) P_2(\cos \theta)$ and  $k(r,\theta) = k_0(r)+k_2(r) P_2(\cos \theta)$. 
The perturbed (unit) fluid flow with rigid rotation and no convection
reads $\vec{u}{\,}^{(1)} = \Omega \partial_\varphi$ and $\vec{u}{\,}^{(2)} = e^{-3\nu/2}(\Omega^2 g_{\varphi \varphi} + 2\Omega K^H_{1t\varphi} + K^H_{2tt}/2)\partial_t$, for a constant $\Omega$.
Hence, the energy density and pressure perturbations only enter 
the second order as  $E^{(2)}(r,\theta) = E_0^{(2)}(r) + E_2^{(2)}(r)P_2(\cos \theta)$, $P^{(2)}(r,\theta) = P_0^{(2)}(r) + P_2^{(2)}(r)P_2(\cos \theta)$.
It is convenient to introduce a rescaled pressure defined by
$\Ptpp_{0/2} := \Ppp_{0/2}/2(E+P)$. 
The first and second order quantities rescale under a perturbation
parameter $\pertp$,
so that any functions $A^{(1)}$ and $A^{(2)}$ 
at first and second order, respectively, enter the model through the scale invariants $\pertp A^{(1)}$ and $\pertp^2 A^{(2)}$.

A convenient substitution to the function $\omega$ in (\ref{fopert_tensor}) is $\tilde{\omega}(r) := \Omega - \omega (r)$, that satisfies the single ODE~(43) in \cite{Hartle1967}. It is integrated from the origin $\of(0):=\tilde{\omega}_c$ outwards. Given that the relevant quantity is $\pertp\tilde{\omega}_c$ one is free to fix either $\tilde{\omega}_c=1$ or $\pertp=1$ (as in \cite{Hartle1967}). For convenience we choose the former. In the exterior,
% \begin{equation}
% \ov (\rv) = \Omega - \frac{2J}{{\rv}^3}, \quad \mbox{ for some constant } J.\label{sol:omega_vac}
% \end{equation}
$\ov (\rv) = \Omega - 2J \rv^{-3}$ for some constant $J$.

The $l=0,2$ sectors of the second-order perturbations can be studied independently.
In the $l=0$ sector a gauge fixing allows to set $k_0=0$ (see \cite{ReinaVera2014}), so that
the only functions involved are $\{m_0,h_0\}$.
In the interior, a first integral (see (64) in \cite{ReinaVera2014}, or (90) in \cite{Hartle1967}) is used to
substitute $h_0^+$ by $\Ptppz$ and the rest of the field equations provide a
inhomogeneous system of first-order ODEs for the set $\{\mf, \Ptppz\}$
(see Eqs.~(61)-(62) in \cite{ReinaVera2014}, or (97), (100) in \cite{Hartle1967}). 
The equations are integrated from a regular origin taking
$\Etppz(0) = 0$ to obtain the perturbed configuration quantities in terms of
$E_c$ (see below). For the exterior we have 
\begin{equation}
\mv(\rv) = \delta M -\frac{J^2}{\rv{}^3}, \;\;\;
\hv(\rv) = \frac{1}{\rv-2M}\left(-\delta M + \frac{J^2}{\rv{}^3}\right),\label{sol:m0ext}
\end{equation}
for some arbitrary constant $\delta M$ \citep{Hartle1967}. 

The $l=2$ sector involves the functions $h_2$, $k_2$ and $m_2$. The field equations  provide a quadrature for $m_2$, a first integral that relates $h_2$ and $\Ptppt$, and a system of coupled first-order ODEs (see Eqs.(67)-(68) in \cite{ReinaVera2014} or (125)-(126) in \cite{Hartle1967}) for the pair $\{h_2, \upsilon := h_2 + k_2\}$. For a regular origin, the set $\{\htwof, \vf\}$ is determined up to one arbitrary constant. The exterior is found in terms of an arbitrary constant $K$ and the Legendre functions of the second kind ($Q_{l=2}^m$) \citep{HartleThorne1968}
\begin{align}
\htwov (\rv) =& K Q^2_2 \left(\frac{r}{M}-1\right) + J^2 \left(\frac{1}{M \rv^3} + \frac{1}{\rv^4} \right),\label{l2ext1} \\
\vf (\rv) =& K \frac{2M}{\sqrt{\rv(\rv-2M)}} Q^1_2 \left(\frac{\rv}{M}-1\right) - \frac{J^2}{\rv^4}.\label{l2ext2}
\end{align}

% % % % % % % % % %% %
\subsection{Tidal problem}
\label{subsection:tidalsetting}
% % % % % % % % % % % %

We summarize next the even sector of the linearized perturbations of a spherically symmetric perfect fluid body due to a quadrupolar tidal field. This problem was analyzed in \cite{Hinderer} using the methods developed in \cite{ThorneCampolattaro} to study nonradial modes of pulsation. For simplicity, we restrict the discussion to the static limit of the perturbations. The (even) tensor perturbation reads in the Regge-Wheeler gauge \citep{Hinderer}
\begin{align}
K_1^T =& \sum_{l,m}
 \left \lbrace   e^{\nu(r)} H_0(r)_{lm} dt^2 + e^{\lambda(r)} H_2(r)_{lm} dr^2 \right.\nonumber\\
&\left. + r^2 K(r)_{lm}(d \theta^2 + \sin^2 \theta d \varphi^2 ) \right \rbrace Y_{lm}(\theta,\varphi). \label{metric_pert_1}
\end{align}
The equations for the different modes $\{l,m\}$ decouple, and those for the $\{l \geq 2, m=0\}$ modes
yield $H_2^+(\rf)_l = H_0^+(\rf)_l$ plus the coupled ODEs (dropping the $m=0$ label)
\begin{align}
&\rf^2 \nuf{}' H^+_0{}_l' = e^{\lf} \left (l(l+1) -2 \right)K_l^+ \label{tidal:H0p}\\
&\qquad + \left(\rf (\lf{}'+\nuf{}') - \left(\rf \nuf{}'\right)^2 -e^{\lf} l(l+1) + 2 \right)H_0^+{}_l, \nonumber\\
&K^+_l{}' = H_0^+{}_l'+ \nuf{}' H_0^+{}_l.\label{tidal:Kp}
\end{align}
The system is integrated from a regular origin, and is usually written as a single second-order ODE for the functions $H_0{}_l$ (see Eqs.~(27)-(29) in \cite{DamourNagar}).

In vacuum, Eqs.~(\ref{tidal:H0p}) and (\ref{tidal:Kp}) hold with $e^{\nuv} = e^{-\lv} = 1-\frac{2M}{\rv}$, and
the general solution for $H_0{}_l$ reads 
\footnote{
  $\hat{P}_l^2 := \left(\frac{2^l}{\sqrt{\pi}} \frac{\Gamma(l+1/2)}{ \Gamma(l-1)}\right)^{-1} P_l^2$ and $\hat{Q}_l^2:= \left(\frac{\sqrt{\pi}}{2^{l+1}} \frac{\Gamma(l+3)}{ \Gamma(l + 3/2)}\right)^{-1} Q_l^2 $.
}
\begin{equation}
H_0^-{}_l(\rv) = a_{lP} \hat{P}_l^2 \left(\frac{\rv}{M}-1\right) + a_{lQ} \hat{Q}_l^2 \left(\frac{\rv}{M}-1\right),\label{solution:H0vacuum}
\end{equation}
for arbitrary constants $a_{lP}$ and $a_{lQ}$ \citep{DamourNagar}.

% % % % % % % % % % % % % % % % % %%%%%%%%
\section{Matching to second order of rotating compact stars}
% % % % % % % % % % % % % % % % % %%%%%%%%

The solutions discussed in the previous section depend on some integration constants left undertermined.
These must be fixed by the relations that the interior and exterior problems satisfy on the common boundary $\Sb$, and are, in turn, related to functions on $\Sb$ that eventually describe the deformation of the surface.
The full set of perturbed matching conditions in this second-order context (\ref{fopert_tensor})-(\ref{sopert_tensor}), from a pure
geometrical description, were consistently derived and discussed in \cite{ReinaVera2014},
using the framework developed by \cite{Mars2005}.
We briefly review the results concerned here before discussing quark stars.

The matching conditions for the first-oder perturbation tensor (\ref{fopert_tensor}) are $\left[\omega \right]= \left[\omega'\right]=0$ (see Proposition 1 and the gauge discussion in \cite{ReinaVera2014}), and therefore
\begin{equation}
J = \frac{1}{6} \ro^4 \of{}'(\ro)\qquad  \Omega =\of(\ro) + \frac{2J}{\ro^3}. \label{JOmega}
\end{equation}
$J$ and $\Omega$ are thus obtained given $\tilde{\omega}_c$ ($=1$). The stellar angular momentum and velocity are $J^S=\pertp J$ and $\Omega^S=\pertp \Omega$. Given some angular velocity $\Omega^S$ as data, $\pertp$ is determined by $\pertp=\Omega^S/\Omega$. The \emph{moment of inertia} is defined as $I := J^S/\Omega^S=J/\Omega$.

For the second order, in the $l=0$ sector we only need for our purposes here (Theorem 1 in \cite{ReinaVera2014})
\begin{align}
&[m_0] = -4 \pi \frac{\ro^3}{M}(\ro-2M) E(\ro) \Ptppz(\ro), \label{m0_matching}\\
& E(\ro)\left(2 \Ptppz(\ro) +\frac{M}{\ro^2} e^{\lambda(\ro)/2}\Xi_0\right)=0. \label{eq:defor_l0}
\end{align}
Given the set $\{\mf(\rf),\Ptppz(\rf)\}$ has been determined in the interior, (\ref{m0_matching})
fixes $\delta M$ in (\ref{sol:m0ext}) as,
 \begin{equation}
  \label{eq:deltaM}
 \delta M = \mf(\ro)+\frac{J^2}{\ro^3} + 4\pi\frac{\ro^3}{M}(\ro-2M)E(\ro) \tilde{P}_0^{(2)}(\ro).
\end{equation}
The total mass, in terms of a fixed $\Eb_c$, reads 
\begin{equation}
M_T(\Eb_c)=M(\Eb_c)+\pertp^2 \delta M(\Eb_c).
\label{eq:total_mass}
\end{equation}
The second order correction to the mass $\delta M^S:=\pertp^2 \delta M$ is usually called the change in mass. The correction to the original Hartle's model comes from the discontinuity (\ref{m0_matching}), yielding the last term in (\ref{eq:deltaM}). Eq.~(\ref{eq:defor_l0}) provides $\Xi_0$ if $\Eb(\ro)\neq 0$ (see below), which describes the star $l=0$ deformation  (in the gauge used) through the average radius of the rotating star $\radius=\ro -\pertp^2 e^{-\lambda(\ro)/2}\Xi_0/2$ \citep{ReinaVera2014}, producing the usual
$\radius=\ro+\pertp^2 \ro(\ro-2M)\Ptppz(\ro)/M$ \citep{Hartle1967}.

The matching conditions for the $l=2$ sector can be split into two sets. The first one contains two purely geometrical (independent of Einstein's equations) relations
\begin{equation}
[h_2] = 0, \quad [k_2] = 0. \label{eq:h2v2}
\end{equation}
The second set is obtained by combining the rest of the geometrical matching conditions with the field equations. That results in the following four matching conditions
\begin{align}
  &\left[h_2'\right]=4\pi \Eb(\ro) \frac{\ro^2}{M}\left\{ h_2(\ro)+ \frac{1}{3} e^{2\lambda(\ro)}\left((\ro-M)^2+M^2\right)\of{}^2(\ro)\right\}\label{eq:l2_a}\\
  &\left[k_2'\right] =4\pi\Eb(\ro)\frac{\ro^2}{M} \left\{h_2(\ro)+\frac{1}{3}\ro(\ro+2M) e^{\lambda(\ro)}\of{}^2(\ro)\right\}\\
  & \left[m_2\right] =\frac{8}{3}\pi \ro^4\Eb(\ro) e^{\lambda(\ro)} \of{}^2(\ro)\label{eq:l2_c} \\  
  &\Eb(\ro)\left\{h_2(\ro)-\frac{1}{4}\nu'(\ro)e^{-\lambda(\ro)/2}
    \Xi_2+\frac{1}{3}\ro^2 e^{\lambda(\ro)}\of{}^2(\ro)\right\}=0. \label{eq:Q2h2}
\end{align}
Again, the last equation provides $\Xi_2$, accounting for the star deformation (eccentricity), if $\Eb(\ro)\neq 0$ (see below).

The two conditions (\ref{eq:h2v2}) fix both the constant from the homogeneous part in $\{h_2^+, \vf\}$ and $K$ from the exterior solution (\ref{l2ext1})-(\ref{l2ext2}). $K$ provides the \emph{quadrupolar moment} $Q^S$ by
\begin{equation}
Q^S =\pertp^2 Q=\pertp^2\left( \frac{8}{5} K M^3 + \frac{J^2}{M}\right).
\end{equation}
A relevant dimensionless quantity independent of $\pertp$ (the rotation) is $\overline{Q}:=Q^S M/J^S{}^2=Q M/J{}^2$ \citep{YagiYunes_waves}.

\begin{figure}
	\centering
	\includegraphics[width=0.9\columnwidth]{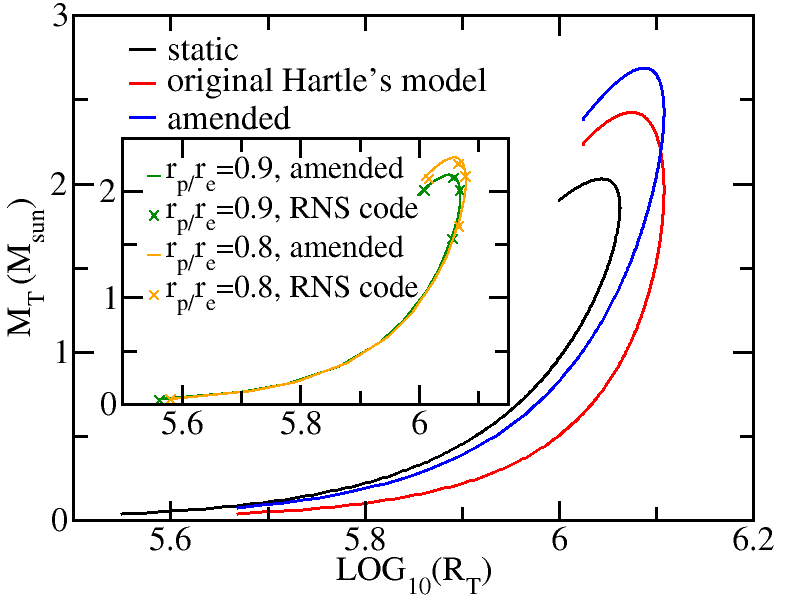}
	\vspace{-0.4cm}
	\caption{Total mass vs radius diagram for strange stars. For the rotating models (red and blue curves) significant differences appear when the total mass is correctly computed. Inset (polar-to-equatorial axis ratio): $r_{p}/r_{e}$=0.9: $\text{L}_{2}$-norm$ \lbrace$amended, Hartle$\rbrace$=$\lbrace0.037,0.075\rbrace$; $r_{p}/r_{e}$=0.8: $\lbrace0.10,0.17\rbrace$.}
	 \label{fig:Colpi1}
\end{figure}

\begin{table}
    \centering
    \begin{tabular}{c c c c c c c  }
        \hline
        $\Eb_c$  & ${\ro}$   & $M$ & $\frac{\pertp^2\delta M^O}{M}$ & $\frac{\pertp^2\delta M}{M}$  &          $\radius $ & ${\lambda}_2 $\\
          $(10^{14}$g\,cm$^{-3})$ & $({R_S})$   & $({M_\odot})$   &    &    &  (km)   &  \\
        \hline
        %(1)
        $4.10 $ &  $31.6$ & $0.038$ & $0.016$  & $0.958$ &$4.67$&$457\times 10^6$\\
        % % % % % % % % % % % % % % % % % %
        $5.66 $ &  $3.42$ & $1.000$ & $ 0.134 $  & $0.624$ &$12.0$&$2540$\\
        % % % % % % % % % % % % % % % % % %
        $6.84 $ &  $2.67$ & $1.400$ & $0.164$  & $0.524$ &$12.7$&$497$\\
        % % % % % % % % % % % % % % % % % %
        $19.18 $ &  $1.85$ & $2.030$ & $0.194$  & $0.310$ &$11.8$&$23$\\
        \hline
    \end{tabular}
    \caption{
    Strange star models.
    $R_S=2M$ stands for the Schwarzschild radii, where $M$ is the static mass.
    %The static radius ($\ro$) is measured in Schwarzschild radii $R_S=2M$, where $M$ is the static mass, while
    %The average perturbed radius
    %and $\radius$ is the average perturbed radius.
    $\pertp^2\delta M^O/M$ is the fractional change in mass computed in the original Hartle's model, while $\pertp^2 \delta M/M$ is the corrected version.
    %${\lambda}_2$ is the dimensionless tidal number. }
    }
    \label{table:colpi}
\end{table}

To illustrate the relevance of the corrections to previous work implied by the above discontinuities
we build numerically equilibrium models of strange stars and compare with Hartle's original approach 
taken by~\cite{ColpiMiller}. We use the MIT bag model with a linear EOS of the type $\Pb = \frac{1}{3}(\Eb - 4B), \;\; \Eb \geq 4B$, where $B$ is the \textit{bag constant}. As in~\cite{ColpiMiller} we  use $B = 56.25$ MeV \,fm$^{-3}$. We consider a range of $E_c$ from $4.1 \times 10^{14}$ to $4.4\times 10^{15}$g cm$^{-3}$. Within this range, the smallest value of $E_c$ generates a non-rotating model of $\sim 0.04M_\odot$, while the largest mass achieved is $\sim 2M_\odot$, corresponding to $E_c = 1.9 \times 10^{15}$ g cm$^{-3}$.

Figure \ref{fig:Colpi1} shows the mass of the configurations against the mean radius for both the static case and the rotating case, assuming a fixed rotational velocity $\Omega^S=\sqrt{M/\ro^3}$ (the mass-shedding limit). The inset shows the comparison with the {\it exact} results, computed with the {\tt rns} code without the slow-rotation approximation~\citep{stergioulas1999keplerian}, for two values of the polar-to-equatorial axis ratio. For a given central energy density the mass increases due to rotation. The correction to the computation of the total mass of the rotating stars leads to significantly higher values than those in \cite{ColpiMiller} (compare red and blue curves). In particular, the maximum mass is $2.69 M_\odot$, $\sim 11$\% larger than that attained in Hartle's approach. In addition, Table \ref{table:colpi} reports the  numerical values of relevant model parameters (compare with Table 1 of~\cite{ColpiMiller}).
 We find that the maximum mass difference is $\sim 0.5 M_{\odot}$, achieved for a density $\Eb_c = 6.34\times 10^{14}\; $ g cm$^{-3}$.

%%%%%%%%%%%%%%%%%
\section{Tidal linearized matching}
%%%%%%%%%%%%%%%%%

A similar analysis can be carried out for the perturbations describing the full tidal problem, Eq.~(\ref{metric_pert_1}), generalising the matching conditions from \cite{ReinaVera2014} to a nonaxisymmetric setup. Such study will be presented elsewhere. However, under the assumption of staticity made in Section \ref{subsection:tidalsetting}, the matching for the axisymmetric $\{l\geq2, m=0\}$ sector of the tidal problem becomes a subcase of the matching given by Eqs.~(\ref{fopert_tensor})-(\ref{sopert_tensor}) after \textit{i)} setting $\omega= 0$ and \textit{ii)} identifying $h(r,\theta) = -H_0{}(r,\theta)/4$, $m(r,\theta) = H_2{}(r,\theta)/4$ and $k(r,\theta) = K(r,\theta)/4$. Proposition 2 in \cite{ReinaVera2014} ensures then that the corresponding spherical-harmonic decomposition coefficients for all $l\geq 2$ satisfy equations equivalent to (\ref{eq:h2v2}) and (\ref{eq:l2_a})-(\ref{eq:l2_c}), leading to
\begin{align}
&[H_0{}_l] = 0, \quad [K_l] = 0, \label{tidal_HK}\\
&\left[H_0{}_l'\right] = \left[K_l'\right] = \frac{4 \pi \ro^2}{M}\Eb(\ro) H_0{}_l(\ro), \quad \left[H_2{}_l\right] =0, \label{tidala}
\end{align}
while the analogous to (\ref{eq:Q2h2}) for all $l\geq 2$ leads to
\begin{equation}
\Eb(\ro) \left(H_0{}_l (\ro)+  \nu'(\ro) e^{-\lambda(\ro)/2} \Xi^{({\rm tid})}_l\right) = 0. \label{tidaldefh}
\end{equation}
Two remarks are in order. First, conditions (\ref{tidal_HK}) are independent of the field equations
and therefore $H_0{}_l$ and $K_l$ will be continuous irrespective of the theory
used. However, the continuity of $H_2{}_l$ (as well as conditions (\ref{tidaldefh})) is a consequence 
of both the geometrical matching plus Einstein's equations for a perfect fluid. For other matter content 
or theory of gravity, the geometric matching conditions from Proposition 2 in \cite{ReinaVera2014} 
must be conveniently combined with the corresponding field equations. Of course, if $\Eb$ presents 
a jump ar $r=\ro$ (and $H_0{}_l(\ro)\neq 0$), so will $H_0{}_l'$ and $K_l'$.

Second, the deformation of the star (in the gauge used) due to the tidal field is encoded in
$\Xi^{({\rm tid})}_l$. It is remarkable that the perturbed matching procedure allows its determination only when 
$\Eb(\ro) \neq 0$, through (\ref{tidaldefh}). This is equivalent to what happens to $\Xi_0$ and $\Xi_2$
in the rotating star setting, 
Eqs.~(\ref{eq:defor_l0}) and (\ref{eq:Q2h2}). However, as shown in \cite{ReinaVera2014}, $\Xi^{({\rm tid})}_l$ 
satisfy the vanishing of the second factor in (\ref{tidaldefh}) even when $\Eb(\ro) = 0$ whenever 
a solution of the problem for all orders of the perturbative expansion exists --this is, in fact, the argument 
implicitly used in the literature, e.g.~\cite{Hartle1967}.

It is convenient to define the function $y := r H_0'/H_0$ in order to compute the Love numbers. The boundary conditions are directly obtained from (\ref{tidal_HK}) and (\ref{tidala}) and read
\begin{equation}
[y] = 4\pi \ro^3 \Eb(\ro)/M.
\label{eq:dif_y}
\end{equation}
This expression recovers the correction addressed in \cite{DamourNagar} for 
homogeneous stars, which is used in \cite{Hinderer_tidal_waves} for
other EOS with nonvanishing energy density at the boundary.
The constant ratio $a_l := a_{lQ}/a_{lP}$ of the exterior solution
(\ref{solution:H0vacuum}) is thus determined from the interior, using (\ref{eq:dif_y}), by
\begin{align}
a_l &= -\left.\frac{\partial_{r_-} \hat{P}_l^2 -  (y_l^-/\ro) \hat{P}_l^2}{\partial_{r_-} \hat{Q}_l^2 -  (y_l^-/\ro) \hat{Q}_l^2}\right|_{\rv=\ro}\nonumber\\
& = -\left.\frac{\partial_{r_+} \hat{P}_l^2 -  (y_l^+/\ro) \hat{P}_l^2 + (4\pi \ro^2 E(\ro) /M) \hat{P}_l^2}{\partial_{r_+} \hat{Q}_l^2 -  (y_l^+/\ro) \hat{Q}_l^2 + (4\pi \ro^2 E(\ro) /M) \hat{Q}_l^2}\right|_{\rf=\ro}. \label{constant_al}
\end{align}
We compare the exterior solution (\ref{solution:H0vacuum}) with the internally and externally generated parts of the gravitational potential $W$ defined in the DSX approach (see section IV.C in \cite{DamourNagar}) in order to relate the constant $a_l$ (\ref{constant_al}) to the \emph{tidal Love numbers} $k_l =\frac{1}{2}\left(\frac{M}{\ro}\right)^{2l+1} a_l $. In the numerical analysis  we concentrate on $l=2$ and we shall use instead the quantity $\lambda_2:=a_2/3$ (see \citet{YagiYunes_waves}).

%%%%%%%%%%%%%%%%%%%%%%%%
\section{Universality of $I$-Love-$Q$ relations}
%%%%%%%%%%%%%%%%%%%%%%%%
  
\begin{figure}
	\centering
	\includegraphics[width=0.9\columnwidth]{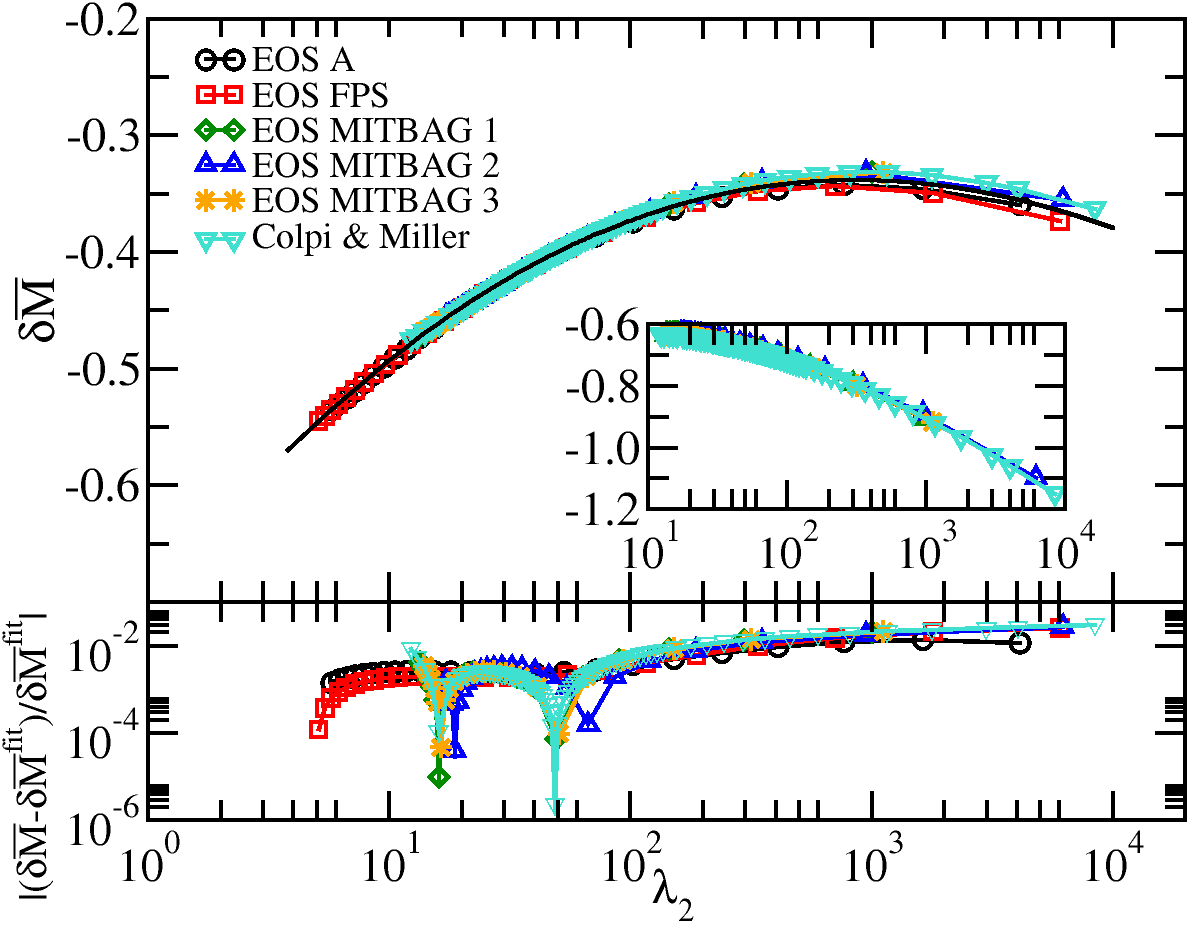}
	\vspace{-0.2cm}
	\caption{Universal relation between $\overline{\delta M}$ and $\lambda_2$ for 6 EOS. The black solid line fits the numerical data to $\log \overline{\delta M}=-0.703 + 0.255 \log \lambda_2 -0.045 (\log \lambda_2)^2 - 5.707 \cdot 10^{-4}(\log \lambda_2)^3 + 2.207\cdot 10^{-4} (\log \lambda_2)^4$. The relative error is shown in the bottom panel. The inset shows the incorrect version of $\overline{\delta M}$ for strange star EOS.}
	 \label{fig:deltaM-k2}
\end{figure}

We turn next to discuss the implications our approach has in the universality of  $I$-Love-$Q$ relations.
Let us first define the rotation-independent (and dimensionless) quantity $\overline{\delta M}:=M^3 \delta M^S/J^S{}^2=M^3 \delta M/J^2$. Using the correct (amended) expressions for $\overline{\delta M}$ and $\lambda_2$ we compute their relation for six different EOS configurations, including two neutron stars and four quark stars. We choose a range of $\lambda_2$ between $10^{0.7}$ and $10^4$, which comprises a range of TOV configurations with mass $M_\odot\leq M \leq 2.3 M_\odot$ (the actual range depends 
on the particular EOS). The numerical results are summarised in Fig.~\ref{fig:deltaM-k2}. We find a strong indication of a universal relation between $\overline{\delta M}$ and $\lambda_2$, as the fit of the numerical data shows (the relative error is displayed in the bottom panel). Such a universal relation is not found for strange stars when using the original ({\it incorrect}) version of $\overline{\delta M}$, as shown in the inset of the top panel.

Therefore, the right use of the perturbed matching yields the corrections used in \citet{YagiYunes_erratum} to find universal $I$-Love-$Q$ relations. Our results show that the second order $l=0$ parameter $\overline{\delta M}$ completes the universal relations that involve the first order parameter $I$, the second order $l=2$ parameter $\overline Q$, and the tidal number $k_2$. This can be used to fix the problems inherent, precisely, to the relations involving (only) the latter three parameters. As discussed in \cite{YagiYunes_waves}, $\overline Q$ is defined from $M$, but the relevant observational quantity is the mass $M_T$ (\ref{eq:total_mass}). From observables one would need to calculate the
corresponding static configuration to find $M$, so as to make the universal $I$-Love-$Q$ relations truly useful in observational astrophysics (at least for weak magnetic fields \citep{Haskell14}). That procedure is model-dependent and, in this regard, the use of $M_T$ instead of $M$ in e.g.~$\overline Q$ is claimed in \cite{YagiYunes_waves} to be of little numerical importance. In this work we have shown that the inclusion of $\overline{\delta M}$ in the $I$-Love-$Q$ relations provides, however, a complete set of relations between all perturbation quantities (to this order), which allows to obtain any such quantity from observational input alone.

\section*{Acknowledgements}
We thank Emanuele Berti for suggesting this investigation and Nikolaos Stergioulas for providing the exact data. Work supported by the Spanish MINECO and FEDER (AYA2013-40979-P, AYA2015-66899-C2-1-P, FIS2014-57956-P), the Generalitat Valenciana (PROMETEOII-2014-069, ACIF/2015/216), and the Basque Government (IT-956-16, POS-2016-1-0075).

\bibliographystyle{mnras}
\bibliography{references}

% Don't change these lines
\bsp	% typesetting comment
\label{lastpage}
\end{document}